\documentclass{ifacconf}

\usepackage{graphicx}      
\usepackage{natbib}        

\usepackage[dvips]{epsfig}    
\usepackage{xcolor}

\usepackage{amssymb}
\usepackage{amsmath}          
\usepackage{enumerate}
\usepackage{tabulary}
\usepackage{enumitem}

{\theoremheaderfont{\upshape} }

\begin{document}
\begin{frontmatter}

\title{Model-free practical PI-Lead control design by ultimate sensitivity principle}

\author[1]{Michael Ruderman}

\address[1]{University of Agder, Department of Engineering Sciences, Norway  \\[2mm]
Correspondence email: \tt \small michael.ruderman@uia.no}

\thanks[*]{On annual (2025) sabbatical at Polytechnic University of Bari \newline 
\textcolor[rgb]{0.00,0.07,1.00}{Post-print of the paper presented at IFAC WC 2026, with corrected missprint of eq. (10); the rest of the paper unchanged}}


\begin{abstract}
Practical design and tuning of feedback controllers has often to get by without a model of the dynamic process at hand. Only some general assumptions about the system dynamics, in this work \emph{type-one stable}, can be available for engineers, for instance in motion control applications and many others. This paper proposes a practical and simple in realization procedure for designing a robust PI-Lead control without modeling. The developed method derives from the ultimate sensitivity principles, known in empirical Ziegler–Nichols tuning of PID controllers, and makes use of some general characteristics of the loop shaping. A three-steps procedure is proposed to determine the integration time constant, control gain, and Lead-element in a way to guarantee a sufficient phase margin, while all steps are served by only experimental monitoring of the output value. Proposed method is demonstrated and discussed with experiments accomplished on a noise-perturbed electro-mechanical actuator system. 
\end{abstract}

\begin{keyword}
Feedback control \sep PID controller \sep Control tuning \sep Model-free design \sep Feedback control synthesis \sep Control parameterization \sep Lead compensator \sep Loop shaping
\end{keyword}

\end{frontmatter}

\section{Introduction}  
\label{sec:1}

Since its first use in \cite{minorsky1922} as three control terms -- proportional, integral, and derivative (PID) -- the PID-controllers became a `working horse', at least in an industrial context, and are considered to be standard feedback strategies in most of the control applications, see also seminal literature e.g. \cite{aastrom2006}. The number of possible generalizations and extensions of PID control, including also nonlinear (\cite{khalil2002,ruderman2025pid}) just to mention here the few, are enormous, equally as of the works dedicated to a PID-control implementation and tuning, see for example \cite{ang2005pid} for overview. Thus, only a little fraction of those, relevant for and in context of the present work, can be mentioned here. Simple analytic PID tuning rules and the associated model reduction were discussed and provided in a seminal work \cite{skogestad2003}. Also in \cite{gyongy2006automatic}, a simple approach to the automatic tuning of PID process-controllers was claimed, while attaining a design-point on the Nyquist diagram. Also the inherent challenges and limitations of an integral feedback action (i.e. in PI and PID control) are well known, for instance associated with the saturation-driven windup effects (\cite{hippe2006windup}) or inability to compensate for the Coulomb friction during motion reversals (\cite{ruderman2025}). Yet, the appealing examples of a relatively recent experimental comparison of the PID auto-tuners (\cite{berner2018experimental}) and the discussions of operational pros and cons of PID-controllers  (e.g. \cite{hagglund2024give}) confirm on top of that a persistent need for the reliable, straightforward, and practically accessible tuning methods.

\begin{figure}[h!]
	\centering \includegraphics[width=0.8\columnwidth]{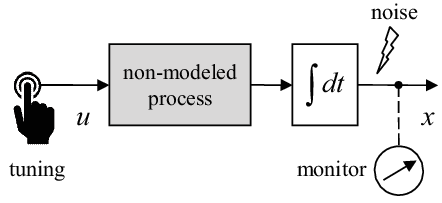}
	\caption{Typical configuration of the model-free controller tuning for type-one dynamic system process. 
}
\label{Fig:1} 
\end{figure}

An automatic (or semi-automatic) tuning of standard feedback controllers (like PID-type), when using only the experimental observations and without knowledge of the system model, was always relevant and in focus of the practicing control engineering, cf. \cite{aastrom2006}. Typical configuration of tuning a (PID) controller in a practical application setting is schematically sketched in Fig. \ref{Fig:1}.
An input-output dynamic process is known to be open-loop stable, but there is no identified system model, or the available modelling assumptions are very vague and general, i.e., without specifying the structure and set of the (determined) parameters. This is a very common scenario when, for example, dealing  with electro-mechanical or hydro-mechanical systems -- a typical industrial equipment for which an already embedded feedback control needs to be appropriately tuned. The 'monitor' implies sensing and processing all necessary data of the process output to be controlled. The 'tuning' involves applying the necessary stimuli to the dynamic process, including closing the loop, and the set and implementation of the tuning rules. An additional assumption that specifies a large class of dynamic processes targeted in the proposed design method is that one deals with a type-one dynamic system, meaning the input-output system response has an integrating behavior, cf. Fig. \ref{Fig:1}. Recall that it renders the system as not stable in the BIBO (bounded-input–bounded-output) sense, cf. e.g. \cite{franklin2019}. Worth noting is also that the type-one systems, i.e. integrating processes, have no open-loop steady-state value and reveals $-90$ deg phase lag at lower frequencies. This can make an experimental tuning of feedback controllers more challenging, cf. \cite{skogestad2003}. Thus, the tuning methodology proposed in this work for type-one processes, can easily be adapted for dynamic processes of type-zero, i.e. without integrating behavior. But it is rather subject for future works, since the type-one is especially relevant for motion control systems and explicitly addressed here.

The rest of the paper is as follows. Section \ref{sec:2} introduces briefly a general set of assumptions for the input-output system plant without requiring an explicit model. The proposed design methodology for PI-Lead control is provided in section \ref{sec:3}. An experimental evaluation shown on a noise-perturbed electro-mechanical actuator system is described in section \ref{sec:4}. Summary and discussion are in section \ref{sec:5}.

\section{Input-Output System Plant}  
\label{sec:2}

We consider an input-output system plant without an explicit model to be available for the control design. The single assumptions, to be known from exploitation of the system, can be summarized as follows.
\begin{enumerate}[label=(\roman*)]
  \item The underlying process is assumed to be SISO (single-input-single-output) and LTI (linear-time-invariant) to a large extent. This implies the linear transfer characteristics dominate over the nonlinear residuals, while the latter cannot destabilize the system through its nonlinear nature. 
  \item The system is of the type-one, i.e. it exhibits a free integrative behavior. This case is typical, for example, for motion control systems where a relative displacement in the generalized coordinates $x$ is the output of interest and to be controlled.  
  \item The system is stable and also minimum phase, in the sense to have no right half-plane poles or zeros. However, an additional process time-delay, i.e. $\exp(-s\tau)$ with an unknown delay constant $0 < \tau < \infty$, is admissible.
  \item The process can be controlled by the input value $u(t)$, while no extra saturations of $u$ are considered for the control design and operation.
  \item The process input and output are assumed to be available in real-time, while the measured output can be inherently affected by an amplitude-bounded unbiased sensor noise. 
\end{enumerate}

Consequently, one can write for the system process
\begin{equation}\label{eq:2:1}
G(s) = \frac{x(s)}{u(s)} = \tilde{G}(s) \frac{1}{s}, 
\end{equation}
where the unknown $\tilde{G}(s)$ satisfies the assumption (iii), and $s$ is the complex Laplace variable. Moreover, the overall process $G(s)$ must have a sufficient phase margin, so that closing the loop $G(s)[1+G(s)]^{-1}$ would not destabilize the output state. If it is not directly given for the open-loop $G(s)$, an additional gaining factor $k > 0$ can be used to allow for a sufficient (in terms of a BIBO stable closed-loop) phase margin of $k G(s)$.

\section{Proposed Design Method}  
\label{sec:3}

The proposed design method is derived from the principles of ultimate sensitivity, see e.g. 
\cite[chapter~4.3]{franklin2019}, for which the phase margin of an open-loop plays the key role. The unknown but stable input-output process is closed by the feedback loop, upon which the (control) gain factor is gradually increased while monitoring the process output variable. For real dynamic processes with the order higher than one, the output exhibits (expectedly) some transient oscillations starting from certain gain value and upwards. With a further fine incrementing of the gain value, one reaches the operation state when the output starts to exhibit permanent oscillations. The corresponding gain is regarded as \emph{ultimate gain}, because if it is further increased, the oscillating output tends to diverge and hence the system destabilizes. For the found ultimate gain value, it is possible to monitor (correspondingly to record) the period of permanent oscillations. The determined, this way, ultimate gain and ultimate period serve as basis for various possible calculations of the P/PI/PID control parameters. The method is known as Ziegler and Nichols tuning rule, cf. e.g. \cite{skogestad2003}, and since its introduction (\cite{ziegler1942}) the approach underwent a variety of adjustments and modifications, see e.g. in \cite{aastrom2006} for details. 

The principle of determining experimentally the ultimate value of a controller gain (we stress that not necessarily a proportional one) and then using the corresponding oscillation frequency to tune the control parameters forms the basis for the design method introduced below. To this end, consider first the standard PI regulator, with the control error $e$, given by the transfer function
\begin{equation}\label{eq:3:1}
C(s) = \frac{u(s)}{e(s)} = K_p \frac{T_i s + 1}{T_i s}, 
\end{equation}
which is parameterized by the control gain $K_p \geq 1$ and the integrator time constant $T_i > 0$. An appropriate tuning of both parameters without a given process model $G(s)$ is known to be a sensitive and not always trivial task.

\subsection{Integrator Time Constant}  
\label{sec:3:sub:1}

Assume first $K_p = 1$ and (without loss of generality) that $G(s)[1+G(s)]^{-1}$ is then stable, cf. section \ref{sec:2}. The latter means that an eventually required gain adjustment by $k$ was achieved so that $G(s)$ is already including $k$ implicitly.  

Since the unknown transfer function $\tilde{G}(s)$ contains only stable poles and zeros and, eventually, a time delay element, its phase response starts from zero at steady-state, i.e. $\angle \tilde{G}(j\omega) \rightarrow 0 $ for $\omega \rightarrow 0$. As by default, $\omega$ is the angular frequency (in rad/sec) and $j$ is the imaginary unit of complex numbers. Following to that, the phase response of an open-loop $C(j\omega) G(j\omega)$ starts at $-180$ deg for $\omega = 0$ and, then, experiences an overall increase by $90$ deg per decade around the corner frequency $\omega^c_{pi} = T_i^{-1}$ of the control $C(j\omega)$. This is under assumption that one starts with a sufficiently high integrator time constant $T_i$ so that a roll-off of $\tilde{G}(j\omega)$ at higher frequencies is not yet effective. 

When gradually reducing $T_i$, the 90 degree phase advance of $C(j\omega)$ will unavoidably confront a phase drop of $\tilde{G}(j\omega)$ at some unknown but higher frequency $\omega > \omega^c_{pi}$. Note that the drop can be smaller, like $-90$ deg per decade, or larger, like $-180$ deg or $-270$ deg per decade, depending on the locus of the dominant poles of $\tilde{G}(s)$. 
\begin{figure}[h!]
\centering \includegraphics[width=0.99\columnwidth]{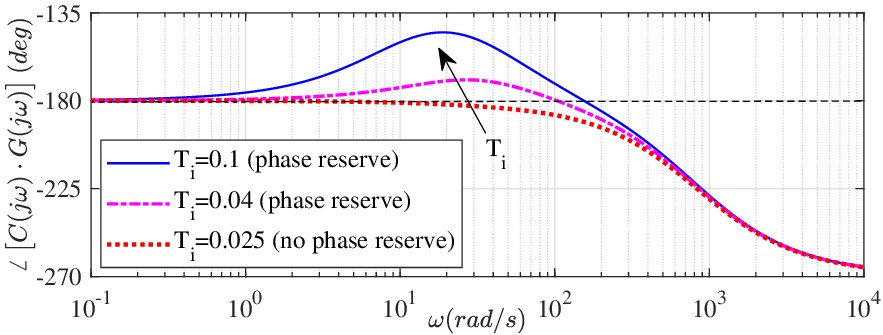}
\caption{Example of the shift and reduction of the phase advance of PI controller $C(j\omega)$ when decreasing the integrator time constant $T_i$.}
\label{Fig:2} 
\end{figure}
Worst case, an even larger phase drop can appear for a dominant time delay. But regardless of the shape of the phase drop, the phase advance of $C(j\omega)$ will be absorbed if further reducing $T_i$, see exemplary illustration of the phase-plot in Fig. \ref{Fig:2}.

As result of a continuously decreasing $T_i$, the output of the closed-loop $C(s) G(s) [1 + C(s) G(s)]^{-1}$ starts to oscillate when applying, for instance, a step reference. By further reducing $T_i$, the step response oscillations become unavoidably less and less damped, due to reduction of the phase margin. Once the permanent oscillations occur, the ultimate integrator time constant $\bar{T}_i$ is determined and so the corresponding ultimate corner frequency $\bar{\omega}^c_{pi} = \bar{T}_i^{-1}$. The situation corresponds to having zero phase margin $\varphi_m = 180 + \angle \bigl[ C(j \omega_{gc}) G(j \omega_{gc})\bigr]$, where $\omega_{gc}$ is the (unknown) gain crossover frequency. The period of the observed, correspondingly recorded, permanent oscillations allows to estimate the gain crossover frequency $\bar{\omega}_{gc}$. Note that whether $\bar{\omega}_{gc} < \bar{\omega}^c_{pi}$ or $\bar{\omega}_{gc} > \bar{\omega}^c_{pi}$ depends on the actual, but unknown, amplitude response of $\tilde{G}(j\omega)$. Yet the largest of both appears crucial for a stable $T_i$ assignment. Following to that, the tuning rule  
\begin{equation}\label{eq:3:2}
T_i = \dfrac{10}{\max \bigl\{ \bar{\omega}_{gc},\, \bar{\omega}^c_{pi}   \bigr\}} 
\end{equation}
is suggested for the integrator time constant. Note that the multiplicative factor ten corresponds to one decade of shifting back (i.e. to the left in frequency range) the phase advance of $C(j\omega)$. The obtained assignment \eqref{eq:3:2} can guarantee that the loop transfer function will have a certain (convex) raising over the $-180$ deg asymptote in the phase response, cf. Fig. \ref{Fig:2}. At the same time, the $T_i$-parameter tuned by \eqref{eq:3:2} is still sufficiently low and, therefore, does not reduce unnecessarily the control bandwidth.

\subsection{Control Gain}  
\label{sec:3:sub:2}

\begin{figure}[h!]
\centering \includegraphics[width=0.99\columnwidth]{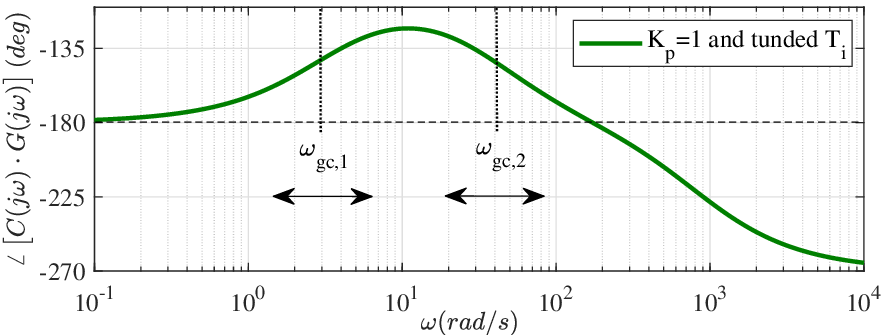}
\caption{Loop phase advance with the phase margin depending on the unknown $\omega_{gc}$.}
\label{Fig:3} 
\end{figure}

The determined above integrator time constant \eqref{eq:3:2} guarantees that the phase response of the open-loop has enough phase advance and does not produce critical oscillations. The initially assigned $\bar{K}_p = 1$ can be, however, either under-tuned or over-tuned depending on the unknown gain characteristics and so cross-over frequency $\omega_{gc}$ of the open-loop, cf. Fig. \ref{Fig:3}.   
By gradually first increasing and then decreasing the control gain, starting from the nominal $\bar{K}_p = 1$ and monitoring the transient overshoot
\begin{equation}\label{eq:3:3}
M = \frac{\max\bigl(x(t)\bigr)-x_{\textrm{ref}}}{x_{\textrm{ref}}}  \quad   \hbox{ for }  \quad K_p \in \bigl[\bar{K}_p \times \delta_{K_p} \bigr] 
\end{equation}
of the process output to the applied step reference $x_{\textrm{ref}}$, one can determine the $K_p$-tuning for which 
\begin{equation}\label{eq:3:3a}
M \in [30,\ldots,40] \hbox{ \%}
\end{equation}
will be achieved. Note that the suggested tuning range $\delta_{K_p} \in [0.1,\ldots,10]$ is also application-dependent and can be further narrowed or expanded around the initial $\bar{K}_p$-value. Further we notice that, independent of the unknown $\tilde{G}(s)$ dynamics, the obtained closed-loop $C(s)G(s)\bigl(1+C(s)G(s)\bigr)^{-1}$ with the already sub-tuned controller \eqref{eq:3:1} will always have a dominant stable pole pair. The latter is conjugate-complex since the controlled response experiences a transient overshoot \eqref{eq:3:3}. Therefore, the damping ratio of that unknown but evident and dominant conjugate-complex pole pair is $0 < \zeta < 1$. Based on the dynamics of a conjugate-complex pole pair, its closed-loop transient overshoot to the step excitation can be computed analytically, cf. e.g. \cite{franklin2019}, by using inverse of
\begin{equation}\label{eq:3:4}
M = \exp\biggl(-\dfrac{\pi \zeta}{\sqrt{1-\zeta^2}} \biggr).
\end{equation}
Also, the corresponding phase margin for the conjugate-complex pole pair is available per calculation as
\begin{equation}\label{eq:3:5}
\varphi_m = \arctan \dfrac{2 \zeta}{\sqrt{ \sqrt{1+4\zeta^4} - 2\zeta^2 }}.
\end{equation} 
Following to \eqref{eq:3:4}, \eqref{eq:3:5}, the obtained control overshoot \eqref{eq:3:3a} will lead to the loop phase margin $\varphi_m \in [40,\ldots,30]$ deg.

\subsection{Lead-based Phase Enhancement}  
\label{sec:3:sub:3}

A Lead-compensator $L(s)$, see e.g. \cite{franklin2019} for basics, is going to be used as a standard flexible tool for the loop shaping, cf. e.g. \cite{messner2007}. This is purposefully  
chosen instead of an (augmented by low-pass filter) pure differential term of a PID control, cf. \cite{aastrom2006}. Note that our goal is to provide an additional phase advance within the critical frequency range around $\omega_{gc}$ and, this way, to enhance the overall robustness and performance of the otherwise unknown loop transfer function $C(s)L(s)G(s)$. This must be achieved without significantly changing the residual frequency characteristics of $C(s)G(s)$. The transfer characteristics of a standard Lead-compensator can be written as
\begin{equation}\label{eq:3:6}
L(s) = K_L \dfrac{\tau s + 1}{\alpha \tau s + 1} \quad  \hbox{ with } \quad 0 < \alpha < 1,
\end{equation} 
and the further design parameters $\tau, \, K_L > 0$. Recall that $\tau$ controls the effective frequency range of a Lead-compensator, while $\alpha$ determines the maximal achievable phase lead $0 < \varphi_L < 90$ deg at the angular frequency
\begin{equation}\label{eq:3:7}
\omega_{\max(\varphi)} = \dfrac{1}{\sqrt{\alpha} \tau}.
\end{equation}   
Also to recognize, from \eqref{eq:3:6}, is that the Lead-compensator gain has $|L(0)| \rightarrow K_L$ and $|L(\infty)| \rightarrow K_L \alpha^{-1}$.

With the above summarized properties in mind, we assign $\alpha = 0.1$. The latter provides a sufficient phase lead $\varphi_L \approx 55$ deg at $\omega_{\max(\varphi)}$ frequency. 
For deciding $\omega_{\max(\varphi)}$ of the Lead-compensator, recall the determined corner frequency $1/T_i$ of the PI controller, and the fact that its phase advance saturates over one decade in frequency range. In order to increase and further extend the overall phase advance, another half-decade of an effective frequency range of the Lead-compensator can also be used. This results in the suggested $\omega_{\max(\varphi)} = 10^{1.5} / T_i$. In order to not affect the overall loop gain at lower frequencies and, thus, not impair the already tuned PI-controller, the Lead-gain is set to $K_L = 1$. Summarizing the above mentioned steps of the Lead-compensator tuning, this results in   
\begin{equation}\label{eq:3:8}
\textcolor[rgb]{0.00,0.07,1.00}{L(s) = \dfrac{ T_i s / 10 + 1}{T_i s / 100  + 1}.}
\end{equation} 

\begin{figure}[h!]
\centering 
\includegraphics[width=0.99\columnwidth]{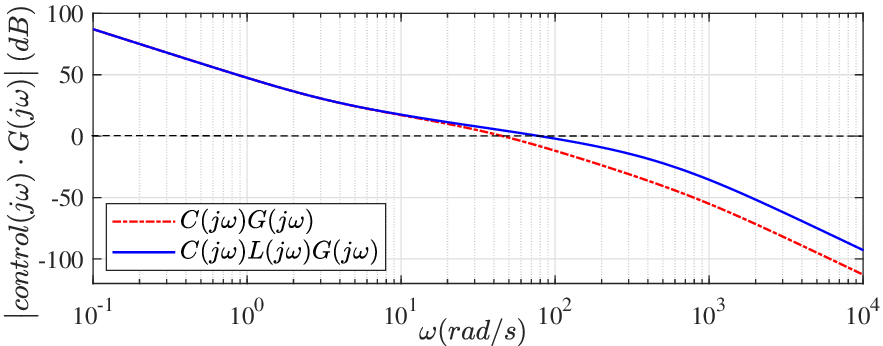}
\includegraphics[width=0.99\columnwidth]{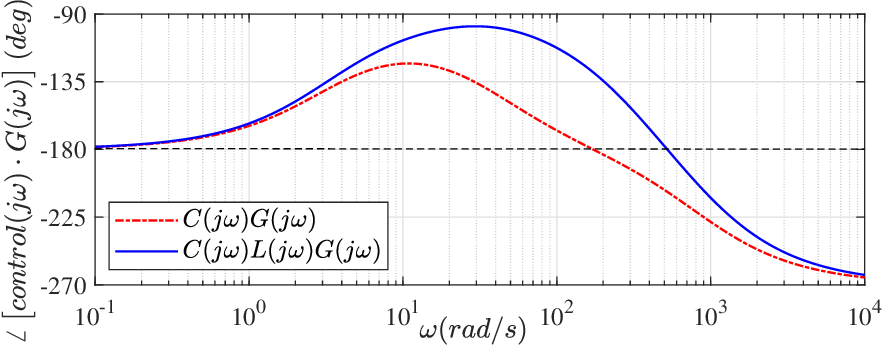}
\caption{Exemplary loop transfer function without ($C \cdot G$) and with ($C \cdot L \cdot G$) the designed Lead-compensator.}
\label{Fig:4} 
\end{figure}

An exemplary exposition of the loop transfer functions without and with the designed Lead-compensator, i.e. $C(j\omega)G(j\omega)$ and $C(j\omega)L(j\omega)G(j\omega)$ respectively, are shown in Fig. \ref{Fig:4}. We emphasize that both the PI-controller $C(s)$ and the Lead-compensator $L(s)$ are designed exactly following the tuning procedure introduced above. The modeled plant transfer function $G(s)$ of the experimental system in use (see \cite{ruderman2022motion,ruderman2025loop}) is only for the sake of the numerical emulation and exposition.

\section{Experimental Evaluation}  
\label{sec:4}

\begin{figure}[h!]
\centering \includegraphics[width=0.7\columnwidth]{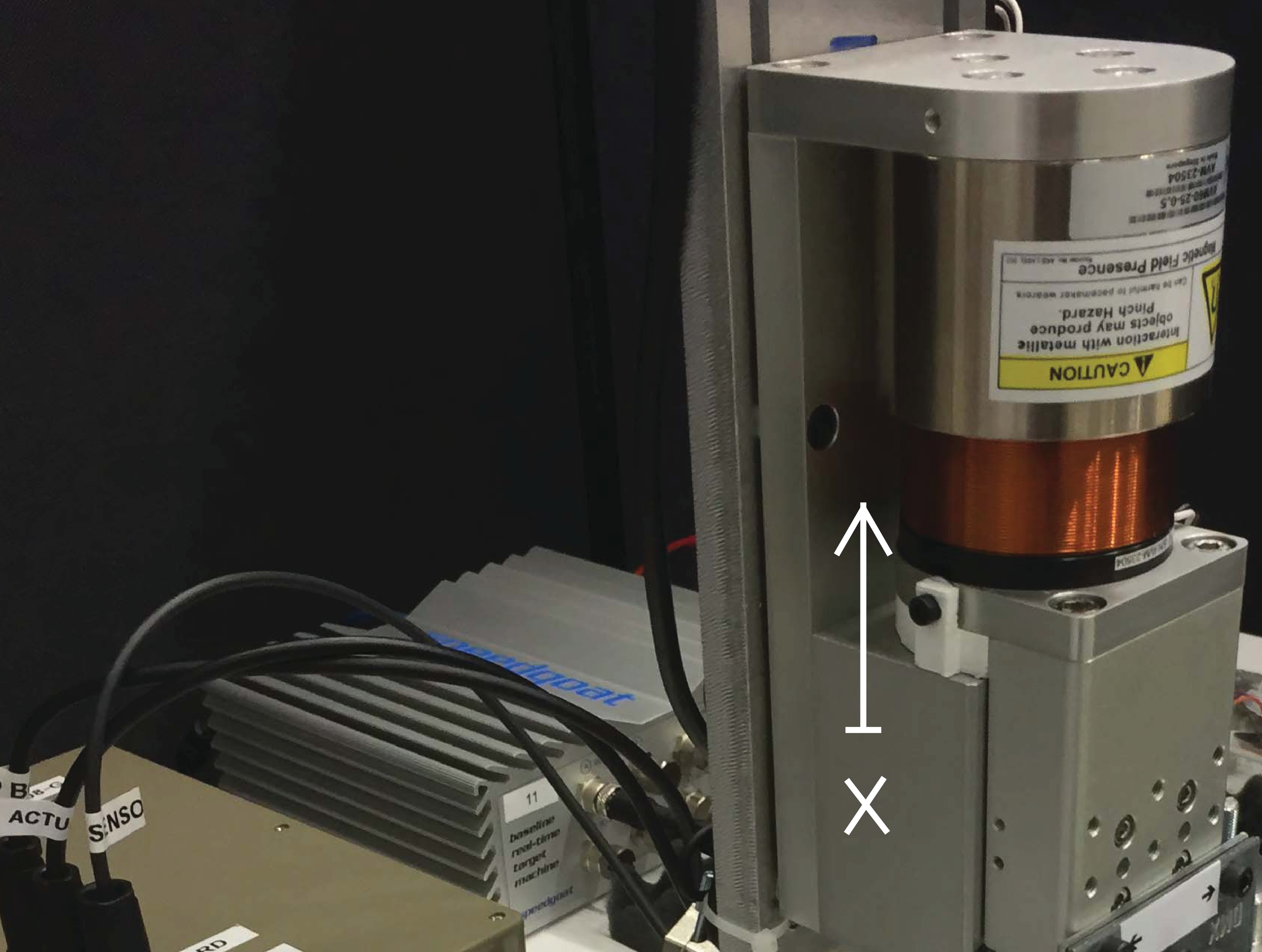}
\caption{Experimental setup of electro-mechanical actuator system (\cite{ruderman2022motion,ruderman2025loop}) with translational degree of freedom $x$, in a laboratory environment.}
\label{Fig:5} 
\end{figure}

The experimental system used in this work for evaluation of the developed practical model-free PI-Lead control design is shown in Fig. \ref{Fig:5}. The system was previously employed in various control related studies, see e.g. \cite{ruderman2022motion,ruderman2025loop}, while more details on modeling and system parameters can be found in \cite{voss2022comparison,ruderman2022motion}. Note that no model or parameter values are used in the following, so that the dynamic process $u(t) \mapsto x(t)$ complies entirely with Fig. \ref{Fig:1} and assumptions made in section \ref{sec:2}. Here $u(t)$ is the controllable input voltage of the voice-coil-motor (VCM) and $x(t)$ in the output relative displacement, measured remotely and essentially noisy. The real-time control board operates at 10 kHz.

Since the acting gravity force is considerable (with respect to the overall control signal range), it is pre-compensated in feed-forwarding by the constant value obtained from the gravity acceleration and the known overall moving mass and input (voltage-to-force) gain. Otherwise, the motion dynamics is additionally perturbed by an input-gain nonlinearity (see \cite{voss2022comparison} for details on the identified system behavior), coil-related force ripples, and nonlinear friction (see \cite{ruderman2023analysis} for basics) in the translational bearing assembly.

The control design is performed by following the tuning steps described in sections \ref{sec:3:sub:1}--\ref{sec:3:sub:3}. It should be noticed that for the closed-loop becomes responsive at all and thus the integrator tuning can start, a loop gaining factor $k=300$ was aleardy initially assigned, cf. section \ref{sec:2}. Thus, the integrator time constant tuning, followed by the successive gain tuning, starts not with $K_p=1$ but with $K_p = k = 300$. With respect to a limited displacement range (about 0.02 m) the step reference here to $x_{\textrm{ref}} = 0.009$ m at this stage. 

\begin{figure}[h!]
\centering \includegraphics[width=0.99\columnwidth]{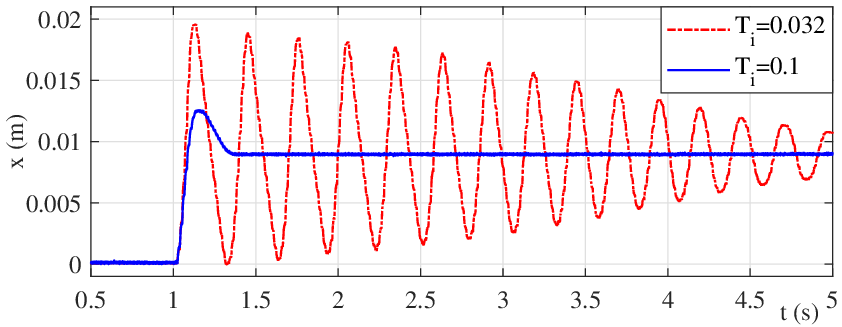}
\caption{Measured step response of PI-control with the initial $T_i=0.1$ and penultimate (i.e. before permanent oscillations appeared) $T_i=0.032$ parameter value.}
\label{Fig:6} 
\end{figure}

The integrator time constant tuning was done by gradually decreasing $T_i$, starting from $T_i=0.1$ and going down until the permanent oscillations of $x(t)$ appeared for $T_i = 0.031$. The measured PI-control response with the initial and penultimate $T_i=0.032$ values are exemplary shown over each other in Fig. \ref{Fig:6}.  
The recorded $\omega_{gc}$ and $\omega^c_{pi}=1/T_i$ values (both in rad/sec), cf. section \ref{sec:3:sub:1}, are exemplary listed in Table \ref{tab:1}.
\begin{table}[h!]
\renewcommand{\arraystretch}{1.5}
    \centering
    \caption{Recorded steps during $T_i$ tuning.}
    \begin{tabulary}{\linewidth}{|C||C|C|C|C|C|C|C|}
    \hline
    $\omega^c_{pi}$ & 10.0 & 20.0    & 25.0      &  28.57  &  30.3  &  31.25  &  \underline{32.26} \\
    \hline 
    $\omega_{gc}$   &  0 & 18.92 & 22.44   &  24.44  &  25.43 &  22.43  &  \underline{20.73} \\
    \hline
    \end{tabulary}
    \label{tab:1}
\end{table}
Note that here $\omega_{gc}$ means the monitored frequency of the appearing oscillations, while $\omega_{gc} = 0$ means no oscillations occurred for the corresponding $\omega^c_{pi}$ control assignment. The last column in Table \ref{tab:1} shows the ultimate values $\bar{\omega}^c_{pi}$ and $\bar{\omega}_{gc}$, cf. \eqref{eq:3:2}. 

\begin{figure}[h!]
\centering \includegraphics[width=0.99\columnwidth]{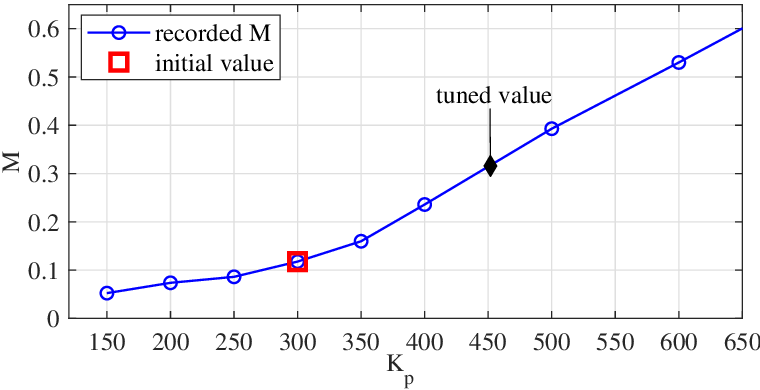}
\caption{Recorded overshoot $M$ over the varying $K_p$ gain.}
\label{Fig:7} 
\end{figure}

For tuning then the control gain, $K_p$ was first incrementally increased and then decreased, starting from the initial $K_p = \bar{K}_p = 300$ value and monitoring the overshoot peaks \eqref{eq:3:3}. The overall recorded $M(K_p)$ dependency is depicted in Fig. \ref{Fig:7} for the sake of a better exposition.   
Both the initial and the determined by the tuning rule \eqref{eq:3:3a} $K_p$-gains are extra highlighted in the figure. The resulting PI-controller, upon both executed above tuning steps, is
\begin{equation}\label{eq:4:1}
C(s) = \frac{139.5 s + 450}{0.31 s}.
\end{equation}

Finally, based on the derived tuning rule \eqref{eq:3:8}, the resulting Lead-compensator is directly given by 
\begin{equation}\label{eq:4:2}
L(s) = \frac{0.031 s + 1}{0.0031 s + 1}.
\end{equation} 

\begin{figure}[h!]
\centering \includegraphics[width=0.99\columnwidth]{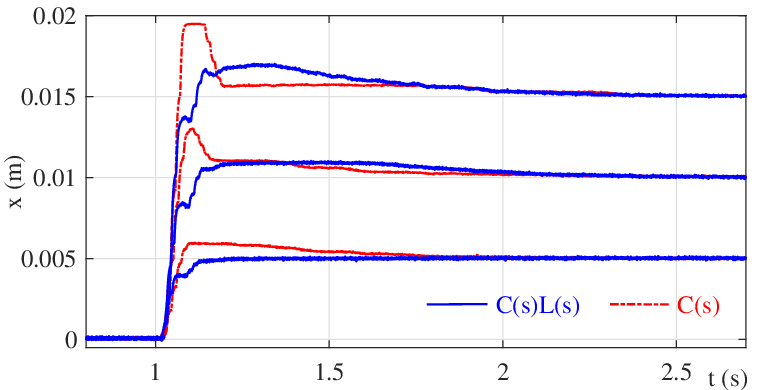}
\caption{Comparison of different step responses of the tuned $C(s)$ and $C(s)L(s)$ controllers.}
\label{Fig:8} 
\end{figure}

\begin{figure}[h!]
\centering \includegraphics[width=0.99\columnwidth]{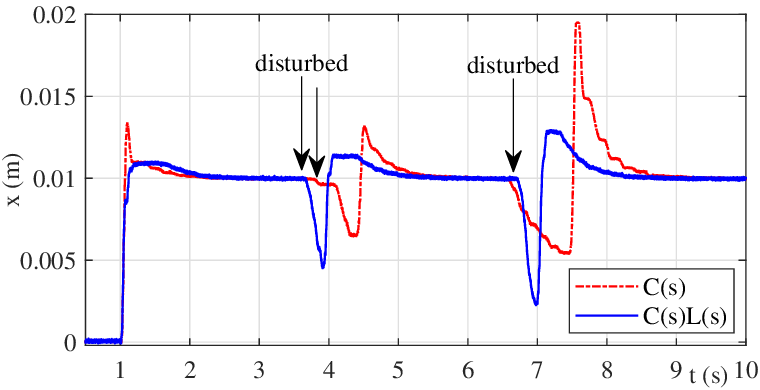}
\caption{Response of the controllers to external disturbance.}
\label{Fig:9} 
\end{figure}

Three reference step values $x_{\textrm{ref}} = \{0.005,\, 0.01,\, 0.015\}$ m are used for comparing the response of the tuned PI and PI-Lead, i.e. $C(s)$ and $C(s)L(s)$, controllers. The experimental results are shown in Fig. \ref{Fig:8}.
Note that the control tuning was performed closer to the second reference point $x_{\textrm{ref}} = 0.01$ m, while the position-dependent impact of nonlinearities, in particular of the actuator input-gain and Coulomb friction, are essentially large. Nevertheless, the performance of the tuned PI- and PI-Lead-controllers are well in accord with analysis of deriving the tuning rules. Recall that especially the nonlinear Coulomb friction cannot be effectively compensated by an integral feedback action during a set-point control task, see \cite{ruderman2025}. The effectiveness and necessity of the Lead-compensator becomes even more evident when applying an external (not necessarily matched) disturbance to the closed-loop system. For that purpose, the moving cartridge of the actuator (cf. Fig. \ref{Fig:5}) was mechanically perturbed by temporary pressing down and then releasing. It should be noted that such manual action cannot be realized with an exactly same effort and same time instants for two consecutive experiments. Nevertheless, the comparison is well interpretable from Fig. \ref{Fig:9} for the $C(s)$ and $C(s)L(s)$ controllers with a step reference $x_{\textrm{ref}} = 0.01$ m. One can recognize the differences in both transient overshoots and settlings after an external disturbance was released.

Finally, the tuned PI-Lead controller is compared with a standard PID controller which was tuned by the Ziegler and Nichols ultimate sensitivity method, cf. e.g. \cite{franklin2019}. Recall that this widely used heuristic tuning method requires to determine the ultimate gain $\tilde{K}_p$ and ultimate period $T_u$ which both correspond to permanent oscillations of the closed-loop system and, thus, characterize a boundary-stable case. Then, the PID control parameters can be assigned as $K_p = 0.6 \, \tilde{K}_p$, $T_i = 0.5 \, T_u$, and $T_d = 0.125 \, T_u$, following the so-called Ziegler and Nichols tuning rules, cf. e.g. \cite{aastrom2006,franklin2019}. Here $T_d$ is the (standard) differentiator time constant of a PID control. The experimentally determined ultimate values are $\tilde{K}_p = 1290$ and $T_u = 0.098$ sec. Following to that, the PID-control tuned by the Ziegler and Nichols method is
\begin{equation}\label{eq:4:3}
C_{\textrm{PID}}(s) = 774 \Bigl(1 + \dfrac{1}{0.049 s}  + 0.012 s \Bigr)  F(s).
\end{equation} 
Recall that an additional low-pass $F(s)$, here designed as Butterworth with a relatively hight cutoff frequency of 1 kHz, is indispensable. It yields $C_{\textrm{PID}}(s)$ to become a proper transfer function and thus implementable. 
\begin{figure}[h!]
\centering \includegraphics[width=0.99\columnwidth]{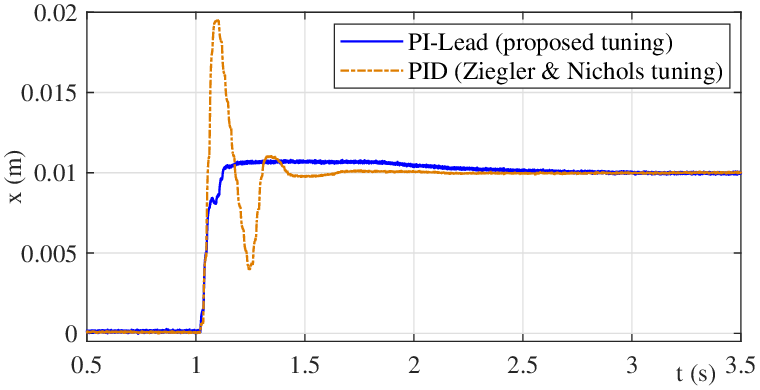}
\centering \includegraphics[width=0.98\columnwidth]{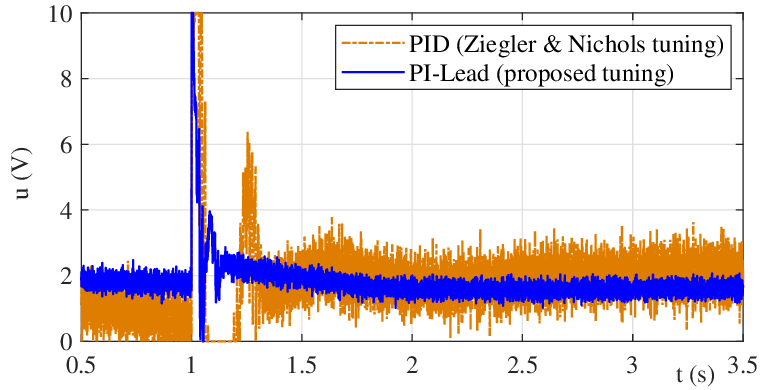}
\caption{Step response of the tuned $C(s)L(s)$ and $C_{\textrm{PID}}(s)$ controllers: measured $x(t)$ above and $u(t)$ below.}
\label{Fig:10} 
\end{figure}
The measured response of the tuned PID control \eqref{eq:4:3} and PI-Lead control \eqref{eq:4:1}, \eqref{eq:4:2} to the step reference $x_{\textrm{ref}} = 0.01$ m are shown opposite each other in Fig. \ref{Fig:10}. One should notice that the final settling of the PID control appears faster since it executes several periods of the transient oscillations and can, this way, compensate faster for the nonlinear Coulomb friction, see \cite{ruderman2025} for details. Also a higher control dither contributes to it, see Fig. \ref{Fig:10} below. Should a PI-Lead controller be re-tuned to have a similar oscillating response, a comparable settling would be achieved. At the same time, one can recognize that the PID control response, tuned by the Ziegler and Nichols method, behaves more aggressive and with very large over- and undershoots, cf. Fig. \ref{Fig:10}. This behavior can be less acceptable in multiple practical applications.

\section{Summary and Discussion}  
\label{sec:5}

A novel practical model-free design of the PI-Lead control, in other words an experimental tuning method, is proposed. The method relies on the ultimate sensitivity principles but differs significantly from the well-known heuristic Ziegler–Nichols ultimate sensitivity approach. The developed and analyzed procedure includes three consecutive steps for (i) determining the integrator time constant, (ii) fixing the control gain, and (iii) assigning the Lead-compensator which enhances the overall robustness and transient performance. All steps are straightforward and simple to implement on the stable dynamical input-output processes of the type-one. The first half of the paper was dedicated to analysis and introduction of the tuning method. Another half of the paper provided a detailed and illustrative experimental study, which is accomplished on a noise-perturbed electro-mechanical actuator system and without use of any modeling or parameters knowledge.  

Following remarks and preliminary conclusions can be stated in a discussion. The proposed tuning method is robust against the feedback noise. Indeed, the proposed tuning of the integrator time constant relies on finding the ultimate oscillations due to the missing phase margin. Such oscillations appear unambiguously (i.e. unmistakably detectable) at lower frequencies and do not depend on the sensor noise. Although the possible nonlinearities and perturbations can reshape oscillations to a certain degree, they do not change principally their period. These properties result directly from both (dominant) poles of the system in the origin: one due to the controller integrator and another due to the type-one system process. Also the monitoring of a transient overshoot, by a subsequent variation of the control gain values, is less sensitive to the system noise. The developed assignment of the Lead-compensator is generic enough to provide a sufficient phase advance in the detected (critical) frequency range of the open-loop and, simultaneously, to not change the already shaped loop characteristics at lower frequencies. Supposedly, the developed tuning method can also be directly used, or at least appropriately adapted, for the processes which include also an oscillatory dynamics, e.g. \cite{ruderman2024adaptive}, time-delay elements, or zero(s) in the origin, e.g. \cite{tavares2021}. These expanding research directions will be subject of the future works.

\section*{Acknowledgement}

The author acknowledges the financial support by NEST (Network for
Energy Sustainable Transition) foundation during the sabbatical 
at Polytechnic University of Bari.

\bibliography{references}             

\end{document}